\documentclass[print%%, %% change to preprint for no formatting
              ]{nsr}

\usepackage{wrapfig}
\usepackage{graphicx,multicol}
\usepackage{amssymb}
\usepackage{amsfonts}
\usepackage{amsmath}
\usepackage{longtable}
\usepackage{rotating}
\usepackage{lscape}
\usepackage{epsfig}
\usepackage{lineno,hyperref}
\usepackage{subfig}
\usepackage{float}
\usepackage{array}
\usepackage{csquotes}
\usepackage{titlesec}
\usepackage{nicefrac, xfrac}

\volume{00}

\artnum{00}

\firstpage{1}

\datesubmitted{XX Month Year}
\datereceived{XX Month Year} 
\daterevised{XX Month Year}
\dateaccepted{XX Month Year} 
\datepublished{XX Month Year}

\doinum{doi/number}

\copyrightyear{2025}

\author[1,2]{A. A. Popov}
\author[1,2]{K. S. Osipenko}
\author[3]{V. I. Scherbakov}
\author[1,2]{A. I. Frank\footnote{Corresponding author e-mail address: \tt frank@jinr.ru}}
\author[1,2]{G. V. Kullin\footnote{Corresponding author e-mail address: \tt kulin@jinr.ru}$^{\orcidlink{0000-0001-9445-211X}}$}
\author[1,2]{M. A. Zakharov$^{\orcidlink{0000-0002-6105-8199}}$}
\author[1,2]{V. A. Kurylev}
\author[3]{D. A. Kolomentseva}

\affil[1]{Joint Institute for Nuclear Research, Joliot-Curie str. 6, 141980, Dubna, Russia}
\affil[2]{Dubna State University, Universitetskaya str. 19, 141980, Dubna, Russia}
\affil[3]{SuperOx, Nauchny proezd 20, 117246 Moscow, Russia}

\runauth{A. A. Popov et al.}

\title{The concept of a superconducting spin flipper – neutron decelerator for a UCN source at a pulsed reactor}

\begin{document}

%% Start line numbering with \begin{linenumbers}, end it with \end{linenumbers}.
%% Or switch it on for the whole article with \linenumbers.
%\linenumbers

\maketitle

%% Insert abstract of the article
\begin{abstract}
\noindent
 The work is devoted to the development of a conceptual design for a gradient spin flipper — neutron decelerator, which is the main component of a designed UCN source for a pulsed reactor. In close cooperation between the JINR group and SuperOx, a preliminary design of a stationary gradient magnet for the adiabatic spin flipper has been developed. A thorough calculation of the magnetic field configuration has been performed. The movement of neutrons in the magnetic field generated by the designed magnetic system has been simulated, and the deceleration time of neutrons in the spin flipper has been analyzed. 
 
The results obtained give grounds for hope that the idea of creating a UCN source based on pulsed accumulation in a trap using non-stationary neutron deceleration is feasible.
\end{abstract}

%% Write your article keywords
\begin{keyword}
ultracold neutrons, UCN source, adiabatic spin flipper, strong magnetic field, flipper-decelerator, birdcage resonator
\end{keyword}

%% Uncomment next two lines if the number of pages is more than 30
%{\hypersetup{linkcolor=black}
%\tableofcontents}

%% Use \section commands to start a section.
\section{Introduction}
%% Labels are used to cross-reference an item using \ref command.
\label{sec:intro}

In the FLNP JINR has begun work on the creation of an ultracold neutron source (UCN) at the IBR-2M reactor \cite{Ananiev77}. This reactor has a repetition frequency of 5 Hz and produces a pulsed flux of thermal neutrons with a pulse width of about 300 microseconds. The pulsed neutron flux density in the reactor moderator is hundreds of times greater than the average, making it very attractive to use the idea of pulsed UCN accumulation in a trap \cite{Antonov69, Shapiro71, Shapiro74}. The latter consists in filling the trap with UCN only during the pulse and effectively isolating it the rest of the time. If during the source operation period the number of neutrons escaping the trap due to various factors is less than the those entering the trap during the pulse, the neutron density in the trap will increase until reaching some equilibrium value. If the neutron losses are not too high, the UCN density in the trap can significantly exceed the time-averaged UCN density in the source.

Unfortunately, in practice, the trap has to be positioned at a considerable distance from the moderator. This necessitates the use of a transport neutron guide. Due to the significant dispersion of the UCN velocities, the spread of flight times in the neutron guide increases rapidly with its length, and even with a length of several meters, the pulse structure of the beam disappears completely. If the variation in flight times exceeds the initial pulse duration but still remains less than the period of their repetition, then the pulse structure can be restored by so-called time focusing or rebunching \cite{Frank00, Arimoto12}. The UCN source based on this idea was proposed in work \cite{Frank22}.

However, an alternative approach to this issue is also feasible. It involves transporting not ultracold, but faster, so-called very cold neutrons (VCN), which are slowed down to UCN energies directly near a trap equipped with a fast valve. In \cite{Nesvizhevsky22}, it was assumed that a rebunching device (a time lens) is positioned on the way from the moderator to the trap, and the neutrons are decelerated by capturing each bunch in a decelerating trap. The speed of latter smoothly changes from the initial VCN velocity of 50 m/s to UCN velocities, after which the neutrons enter the main trap.

Somewhat later, it was realized that if, in contrast to \cite{Nesvizhevsky22}, neutrons are slowed down not by smoothly changing their velocity, but by relatively rapidly changing the energy of all neutrons by the same amount, then the range of neutron velocities that turn into ultracold ones after such a deceleration decreases with increasing energy taken away and can be relatively small. Thus, the VCN flux just before the decelerating device, and the UCN flux at the trap entrance, can remain a pulsed structure even when VCN is transported to a considerable distance \cite{FrankPEPANL23}. This idea is the basis for a new concept of a UCN source with pulsed filling of the trap \cite{FrankArXiv24}.

To decelerate neutrons a local magnetic resonance device, a gradient or adiabatic spin flipper, is proposed to be used, which will slow down very cold neutrons to UCN energy, just before entering the trap. This paper describes this device and its physical properties.

\section{Operation principle of the flipper-decelerator}
%% Labels are used to cross-reference an item using \ref command.
\label{sec:sec2}
The neutron deceleration method based on a nonstationary spin flip in a magnetic field was first proposed in 1960 by G.M. Drabkin and R.A. Zhitnikov \cite{Drabkin60} as a method for producing "supercold" neutrons. This work appears to be the first to draw attention to the fact that seems obvious today, that with this spin flip method, the neutron energy changes. A similar idea was discussed by Kruger twenty years later \cite{Kruger80}. Soon after, the change in neutron energy during a resonant spin flip was first observed experimentally \cite{Alefeld81}, which associated certain difficulties at that time.
In the middle of the last century, resonant spin-flippers based on the combined action of permanent and orthogonal alternating magnetic fields with a fixed frequency became widespread. For an effective spin flip, it is necessary that the neutron be in the region of these two fields for a well-defined time. Moreover, it is necessary that the area of action of the two fields is sufficiently well limited. This condition is easily met for beams of thermal and cold neutrons with a not too wide range of velocities. However, it turns out to be impossible in the case of UCN. The development of work with the latter required the creation of a so-called gradient or adiabatic flipper \cite{Egorov74, Luschikov84, Grigoriev97, Holley12}.

The action of such a spin-flipper is based on a combination of two magnetic fields: a permanent but depending on the coordinate field $B(z)$ directed along the $Z$ axis, and an orthogonal field $B_1$ oscillating with a frequency of $f$ or rotating with a frequency of $\Omega = 2\pi f$. Below we will proceed from the case of a rotating field, since an oscillating field can be represented as the sum of two fields rotating in opposite directions, of which we will be interested in only one. It is convenient to explain the dynamics of the magnetic moment of the neutron and the spin associated with it using a coordinate system rotating with a frequency $\Omega$ in which field $B_1$ is constant. It is known that analyzing the behavior of the magnetic moment of a neutron and its associated spin in a rotating coordinate system, it is necessary to formally add a fictitious magnetic field  $B_{\Omega}=-{\hbar\Omega}/{2\mu}$ to the physical fields and analyze this combination of fields.

In the case under consideration, in the rotating coordinate system, we are dealing with the field ${\widetilde{B}}_z=B(z)-B_{\Omega}$, directed along the $Z$ axis and the constant field $B_1$ orthogonal to it. The magnitude and direction of the total field $B_{eff}={\widetilde{B}}_z+B_1$ depends on the $z$ coordinate, and if the neutron is moving along the $Z$ axis, then on the time of movement. As the neutron moves in the decreasing field $B(z)$, the total effective field $\mathbf{B}_{eff}$ decreases in magnitude and changes its direction, taking the value $\mathbf{B}_{eff}=\mathbf{B}_1$ in the resonance region. At further movement $z$-component of the field $\mathbf{B}_{eff}$ becomes negative and its absolute value increases again. At a considerable distance from the resonance point, where $B_{eff}\gg B_1$ the field $\mathbf{B}_{eff}$ turns out to be directed in the direction opposite to the true field $\mathbf{B}(z)$.

If the change in the field parameters occurs slowly enough, then the magnetic moment, precessing around the time-varying field $\mathbf{B}_{eff}$, generally "follows" its direction, so that the angle between the direction of this field and the magnetic moment remains small. Therefore, after a significant distance from the region of resonance, the magnetic moment, following the field $\mathbf{B}_{eff}$, turns out to be oriented in the opposite direction to the initial one.

The quantitative criterion for the smallness of the angle of the magnetic moment deviation from the field vector, that is, the adiabaticity condition, is the ratio
\begin{linenomath}
\begin{equation}
   K=\frac{\omega_1}{\dot{\omega}}\gg1,                                                                                                   
\label{eq:eq1}
\end{equation}
\end{linenomath}
where $\omega_1 = \gamma B_1$ and $\dot{\omega}=\gamma{\dot{B}}_{eff}$, and $\gamma$ is the gyromagnetic ratio for the neutron. In the resonance region, where  $B_{eff}\approx B_1$,  the adiabaticity condition takes the form:
\begin{linenomath}
\begin{equation}
   K=\frac{\gamma\ B_1^2}{v\left(dB/dz\right)}\gg1,                                                                                                   
\label{eq:eq2}
\end{equation}
\end{linenomath}
where $v$ is a neutron velocity in the resonance region. In practice, the adiabaticity condition \eqref{eq:eq2} is not very strict, and a satisfactory spin flip efficiency can be obtained already for the values $K\approx3\div4$.

Possible doubts about the non-stationary nature of physical processes in a gradient flipper were removed by an experiment \cite{Weinfurter88}, in which it was demonstrated that a spin flip in such a device actually results in a change in the neutron energy in the same way as in the well-known resonant flipper \cite{Alefeld81}.

\section{Preliminary requirements for the magnetic system of a gradient flipper — neutron decelerator}
%% Labels are used to cross-reference an item using \ref command.
\label{sec:sec3}
When determining the requirements for the magnetic system of a flipper-decelerator, it is necessary to remember that a beam of neutrons must pass through the magnetic field formed by it. Therefore, it seems natural to assume that the magnetic system should have cylindrical symmetry, that is, it will be a high-current solenoid. Such a geometry was proposed in work \cite{FrankArXiv24}. In this work to form a high-frequency rotating field it was proposed to use a "birdcage" type resonator, which also has cylindrical symmetry and does not contain matter in the neutron path.

The magnetic system must be such as to ensure high efficiency of the spin flip with the maximum possible amount of energy transferred. At the same time, the neutron deceleration time and especially the dispersion of these times should be as small as possible. It follows that the areas of rise and fall of the field should be as short as possible, and for the adiabaticity parameter \eqref{eq:eq2} to be large enough, they should be separated by an area with a relatively small field gradient.

Entering the region of the increasing field, where the field gradient is large, the neutron slows down, decreasing its kinetic energy by value of ${\Delta E}(z)=-\mu B(z)$. In the region of a slowly changing field with a relatively small field gradient, it is affected by the high-frequency field $B_1$, rotating with a frequency of $\omega_1$, thus creating conditions for an adiabatic spin flip. Here, as noted above, the total neutron energy changes by value of ${\Delta}E_{rez}=2\mu B=\hbar\omega_1$. After passing through the spin flip region, the neutron enters the region of rapid decrease in field strength, where it slows down to a speed typical of the UCN.

If the flipper-decelerator is designed correctly, then the UCN output from it retains to one degree or another the pulsed flux structure that occurred at the entrance to this device. The idea of pulsed accumulation of UCN involves the presence of a fast valve at the entrance to the trap that allows neutrons to pass through for a short time when a bunch of neutrons enters the trap and locks it until the next bunch arrives. Creating a mechanical valve that meets the necessary requirements is a rather difficult task. Therefore, in \cite{FrankArXiv24}, it was proposed to take advantage of the fact that the UCN at the output of the flipper moderator is polarized and if polarized neutrons are also stored in the trap, a region with a strong magnetic field can be used to lock them. The latter can be the field of the flipper decelerator, and to ensure the pulsed nature of the trap filling, the system can be supplemented with a second flipper, the purpose of which is to prepare neutrons with the desired polarization sign. Technically, it is simply possible to supplement the design of the flipper decelerator with another area with a rotating high-frequency field located in the region of the decreasing stationary field. This additional flipper would only operate for a short time, equal to the time it takes for the UCN bunch to pass through it. At the same time, strictly speaking, it is not actually a valve but only prepares a given spin state of neutrons. As a result, polarized neutrons will be stored in the trap, for which the main magnetic field is a barrier throughout time when the flipper is turned off. Note that when the flipper is turned on, it inverts the spin of both neutrons entering and exiting the trap. After the spin is flipped, the magnetic field becomes retracting for them, so during the operation of the flipper, an inevitable outflow of neutrons from the trap occurs. However, this circumstance is typical for the pulsed trap filling method and is accounted when calculating the neutron density inside the trap \cite{Shapiro71, Shapiro74}.

In relation to the above, several regions can be identified in the magnetic field profile: an entry and primary deceleration region characterized by a large field gradient, a region with a relatively small and constant gradient where spin flip occurs, a region of further deceleration with a sharp decrease in field strength, and a region of residual field containing an area where secondary spin flip is occur (see Fig. \ref{fig:fig4} below).

The purpose of each of these regions defines the basic requirements for the configuration of the magnetic field. Let's briefly focus on the considerations that form the basis for the draft design of the device described.

First, let us recall that the purpose of the device is to decelerate neutrons with as little variation in deceleration times as possible, which inevitably increases the bunch duration at the entrance to the trap. The transit time through the section with an inhomogeneous magnetic field $B(z)$ is
\begin{linenomath}
\begin{equation}
   t=-\frac{m}{\mu}\int^{z_2}_{z_1}{\frac{dV/dz}{dB/dz}dz},                                                                                                   
\label{eq:eq3}
\end{equation}
\end{linenomath}
where 
\begin{linenomath}
\begin{equation}
   V(z)=V_0\sqrt{1-\frac{2\mu B(z)}{mV_0^2}},                                                                                                   
\label{eq:eq4}
\end{equation}
\end{linenomath}
$V_0$ and $V(z)$ are the neutron velocities in the absence of a field and in the field region, respectively.

From eq.~\eqref{eq:eq3} it can be seen that in order to minimize the flight time for a region with a varying field, it is necessary to have the largest possible field gradient. However, within the region of initial deceleration, the neutron loses approximately half of its initial energy, and the velocity ratio  $\chi(z)=V(z)/V_0$   changes relatively slightly during its passage through this region. At the same time, towards the end of the deceleration region, the neutron loses a significant part of its energy, which implies that within this region, the value of $\chi(z)$ varies from a value order of $1/\sqrt{2}$ to zero, which makes a significantly greater contribution to the integral of eq.~\eqref{eq:eq3} compared to the first half of the path. Consequently, the field gradient requirements in this region are more stringent compared to those in the primary deceleration region.

Let us now consider the spin flip region and estimate the basic requirements for the field parameters that served as the initial input parameters for the design of the flipper magnetic system. Assuming $K>4$ for desired coefficient of adiabaticity, we obtain from eq.~\eqref{eq:eq2} a relationship between the strength of the rotating field and the gradient of the stationary field
\begin{linenomath}
\begin{equation}
   \frac{dB}{dz}<\frac{\gamma B_1^2}{4V},                                                                                                   
\label{eq:eq5}
\end{equation}
\end{linenomath}
where $V$ is the neutron velocity in place of spin flip. To estimate it, we assume that the magnetic field strength of the flipper is of the order of 20 T, which is practically achievable under modern conditions, and the UCN energy after deceleration by the flipper is $E_{UCN}=150$ neV. This means that the energy of the neutrons entering the flipper-decelerator should be $E_{VCN}=2.55$ $\mu$eV, and in place of spin flip should be 1.35 $\mu$eV. This energy corresponds to a velocity of approximately 16 м/с. Substituting the numerical values for the velocity and gyromagnetic ratio of the neutron $\gamma=1.82\times{10}^8$ T$^{-1}$c$^{-1}$ into eq.~\eqref{eq:eq5}, we can derive the condition for the required permanent magnetic field gradient  $\frac{dB}{dz}<2.8\times{10}^6\cdot B_1^2$. The choice of the high-frequency field strength magnitude is somewhat arbitrary, and the limiting factor is the power output of the corresponding radio frequency (RF) system. Based on our estimates, it seems reasonable to set $B_1\approx 1$ mT in our calculations, which leads to the following condition $\frac{dB}{dz}<2.8$ T/m.

Regarding the determination of the parameters of the magnetic fields that ensure the adiabatic condition for the flipper-valve, these depend on the selection of the field strength $B_0$ at its location. Assuming that the flipper-valve is located within the region of the residual magnetic field of the main system, for instance, in a field on the order of 0.5 T, we obtain that the velocity of neutrons in this region is at least 2.5 times lower than at the position of the main flipper. Accordingly, the requirements for the ratio between the gradient of the constant field and the amplitude of the radiofrequency (RF) field  are reduced proportionally.

An essential aspect in the design of a magnetic field system of the flipper-decelerator is the extent of the spin flip region, which, to a rough approximation, determines the size of the region where adiabatic conditions need to be satisfied. Obviously, where $B_{eff}\gg B_1$, the effective field is directed almost along the $z$ axis, and its main rotation occurs in the region where $\left|B_{eff}\right|$ not too different from $B_1$. Since, when the adiabaticity condition is met, the neutron spin practically "follows" the direction of the field $B_{eff}$, the region of significant rotation of the vector $\mathbf{B}_{eff}$ determines the region of spin flip. To roughly estimate its extent, we assume that the main spin rotation occurs in a region where $\left|B_{eff}\right|<10B_1$ that leads to an estimation of its length $\ell\approx20B_1\left(\frac{dB}{dz}\right)^{-1}$,  that is, the order of $\ell\approx0.7$ см in the region where the main flipper is located. 

Considering that a neutron beam with a certain effective transverse dimension needs to pass through the flipper, it is necessary to ensure that the conditions for efficient neutron spin flip in the entire beam section are met. This necessitates certain requirements regarding the radial uniformity of the field B0 and the length of the region where the resonant conditions hold. For the magnitude of radial field inhomogeneity, a condition has been adopted whereby the deviation $\Delta B(r)$ in the magnitude of the constant field $В$ from its value along the beam axis must satisfy the following condition ${\Delta B}(r)\leq\ell\left({dB}/{dz}\right)$. Obviously, the presence of radial inhomogeneity necessitates the precise fulfillment of the resonant condition $\mathbf{B}_{eff}=\mathbf{B}_1$ not in a single section of the beam but in a cylindrical volume, whose length dependents on the maximum value of radial inhomogeneity. This region's length is evidently $L={\Delta B_{max}}/\left({dB/dz}\right)$, therefore, the adiabaticity condition should be satisfied throughout the entire region with a length $\ell+L$.

 \section{A conceptual design for a superconducting magnetic system}
\begin{figure}[!htb]%% placement specifier
%% Use \includegraphics command to insert graphic files. Place graphics files in working directory.
\centering%% For centre alignment of image.
\includegraphics[width=0.7\linewidth]{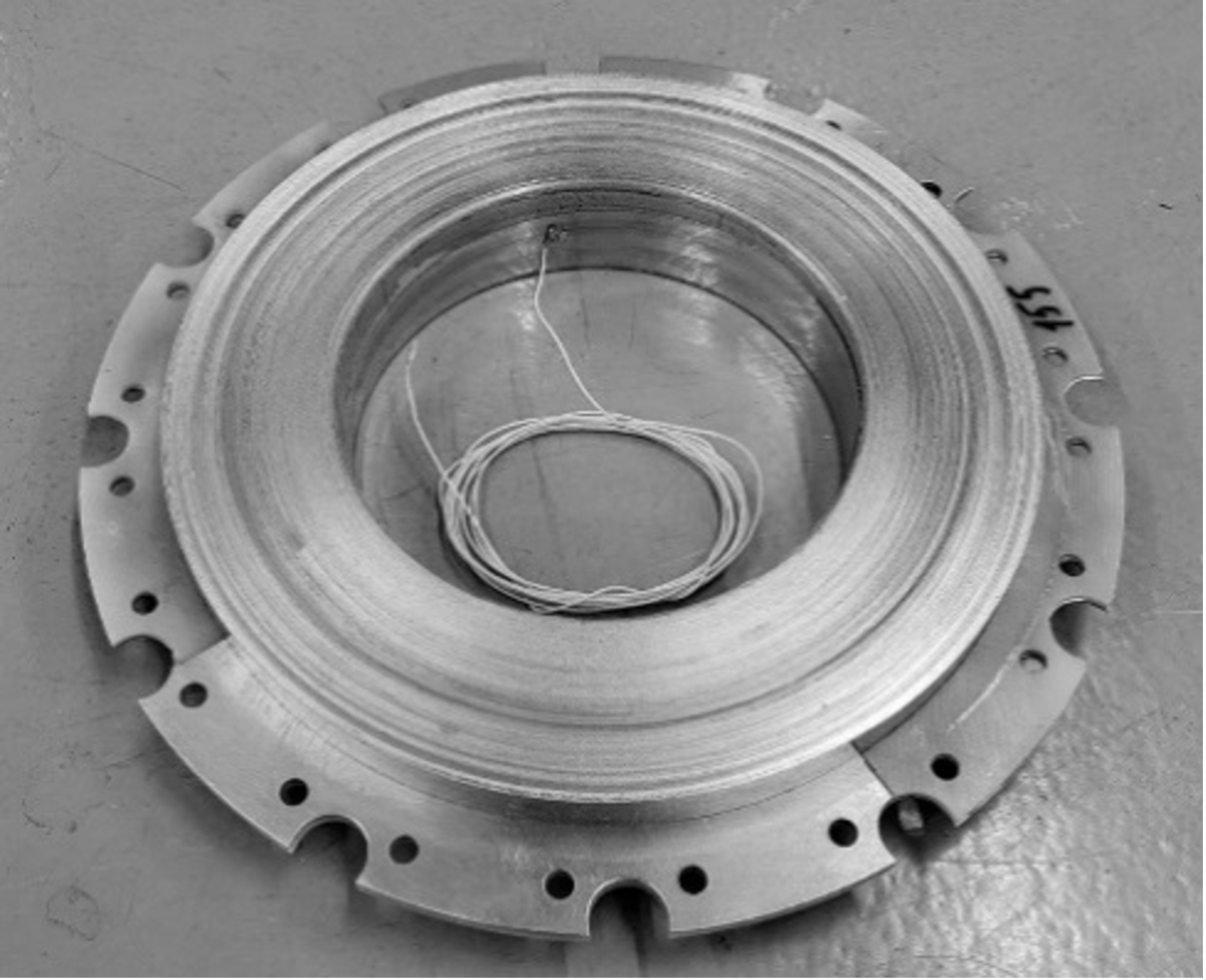}
%% Use \caption command for figure caption and label.
\caption{A prototype of a pancake made from HTS tape.}\label{fig:fig1}
\end{figure}

%% Labels are used to cross-reference an item using \ref command.
\label{sec:sec4}
The main element of the flipper magnetic system is a solenoid, which consists of a set of pancakes (a flat, double coil) manufactured from high-temperature superconducting (HTS) tape of the 2nd generation. The field characteristics of this material are well-known and have been described in \cite{Molodyk21}. In 2022, a magnet was produced and tested that was entirely composed of 2nd-generation HTS tape \cite{Baburin24}. An example of a prototype of a pancake with an internal diameter of 108 mm is shown in Figure~\ref{fig:fig1}.

\begin{figure}[!htb]%% placement specifier
%% Use \includegraphics command to insert graphic files. Place graphics files in working directory.
\centering%% For centre alignment of image.
\includegraphics[width=\linewidth]{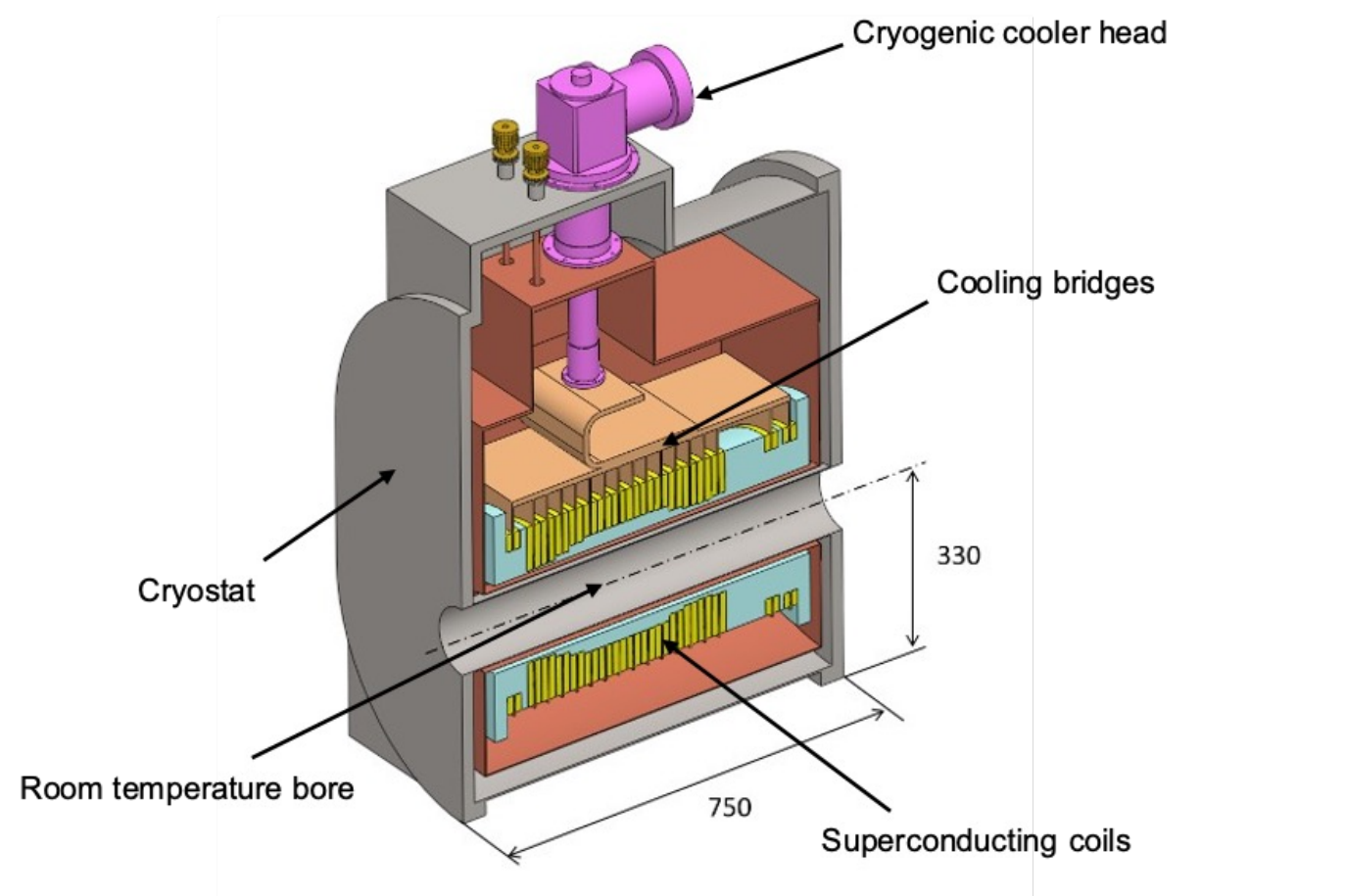}
%% Use \caption command for figure caption and label.
\caption{Design of the magnetic system and cryostat.}\label{fig:fig2}
\end{figure}

\begin{figure}[!htb]%% placement specifier
%% Use \includegraphics command to insert graphic files. Place graphics files in working directory.
\centering%% For centre alignment of image.
\includegraphics[width=\linewidth]{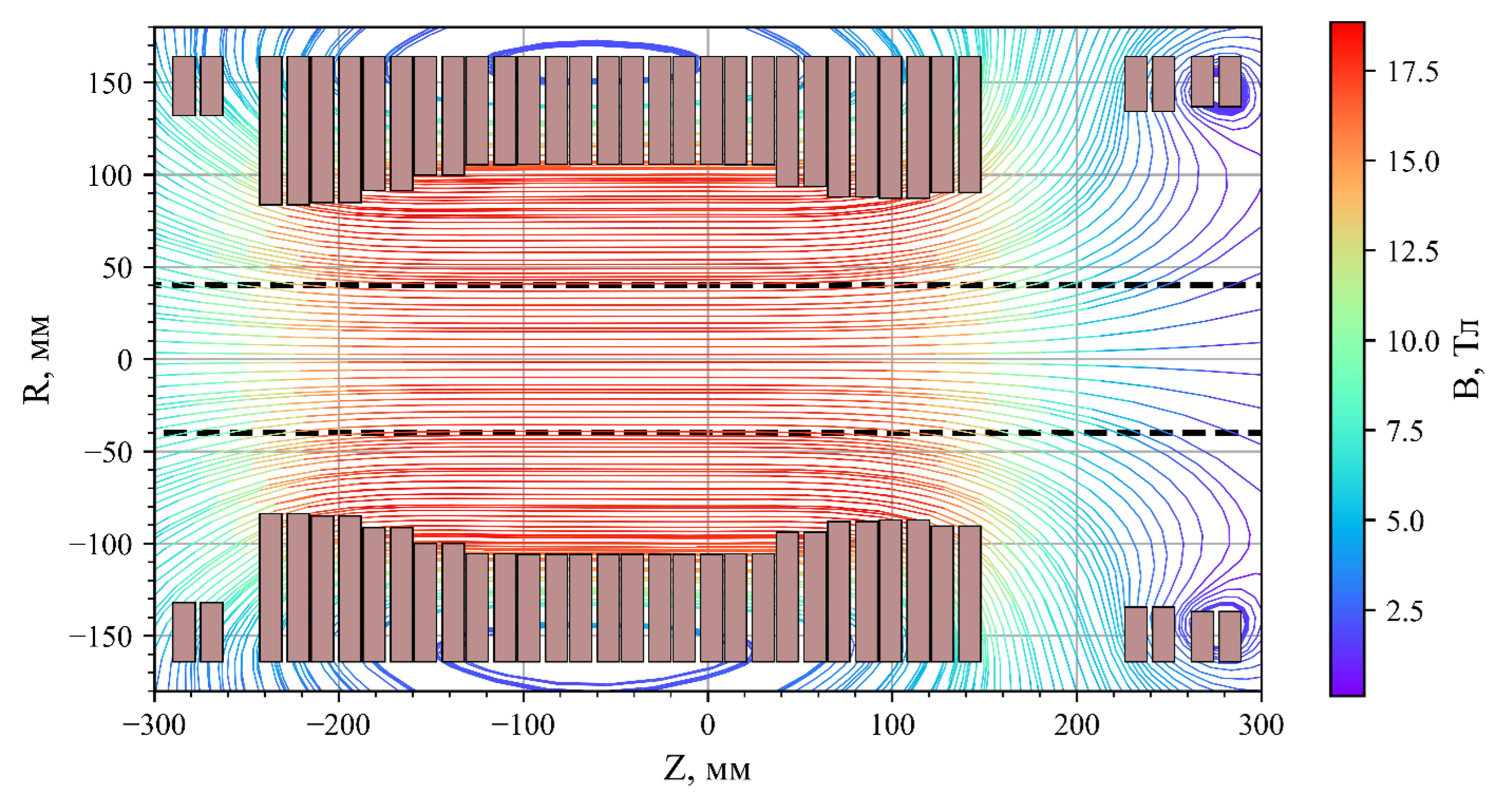}
%% Use \caption command for figure caption and label.
\caption{The configuration of the supply coils and the magnetic flux density distribution. The dashed line indicates the region enclosed by the neutron guide.}\label{fig:fig3}
\end{figure}

Figure~\ref{fig:fig2} shows the design of the flipper magnetic system and cryostat. The magnetic system and magnetic flux density distribution are shown in Figure~\ref{fig:fig3}. A required magnetic field configuration is generated by current flowing in double HTS coils with different radial dimensions. Fourteen coils are the main ones. They create a field whose strength reaches 18T at the center of the solenoid. Three auxiliary coils, one for input and two for output, are used for the field correction. These coils help to achieve the desired field profile in the entry region and in the main deceleration region. Two correction coils at the exit of the solenoid operate with reversed current, generating a small field to reduce the residual field from the main coils. This leads to a sharper field decline in the main deceleration region, reducing neutron deceleration time and its dispersion.

\begin{figure}[!htb]%% placement specifier
%% Use \includegraphics command to insert graphic files. Place graphics files in working directory.
\centering%% For centre alignment of image.
\includegraphics[width=0.8\linewidth]{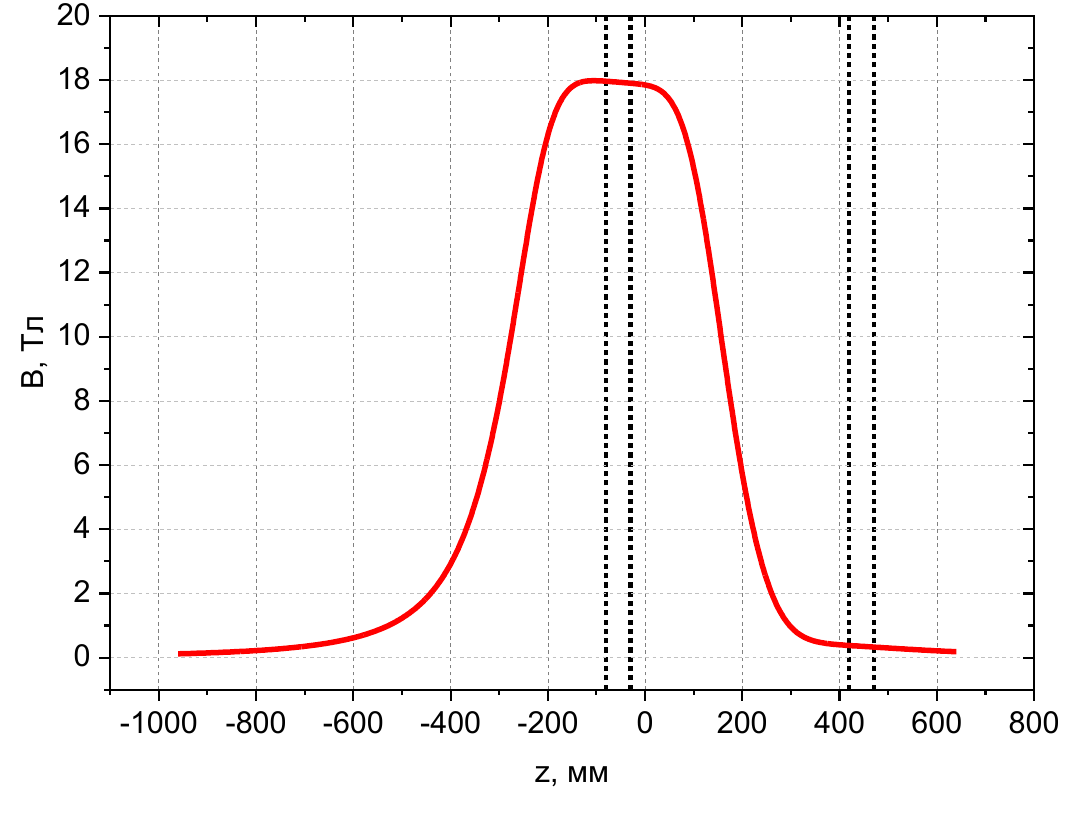}
%% Use \caption command for figure caption and label.
\caption{Magnetic field profile along the solenoid axis.}\label{fig:fig4}
\end{figure}

Figure~\ref{fig:fig4} shows the magnetic field distribution of a magnet. Dashed lines indicate the boundaries of two spin flip regions. 

\begin{figure}[!htb]%% placement specifier
    % Use \includegraphics command to insert graphic files. Place graphics files in working directory.
    \centering %% For centre alignment of image.
    \hfill
    \includegraphics[width=0.495\linewidth]{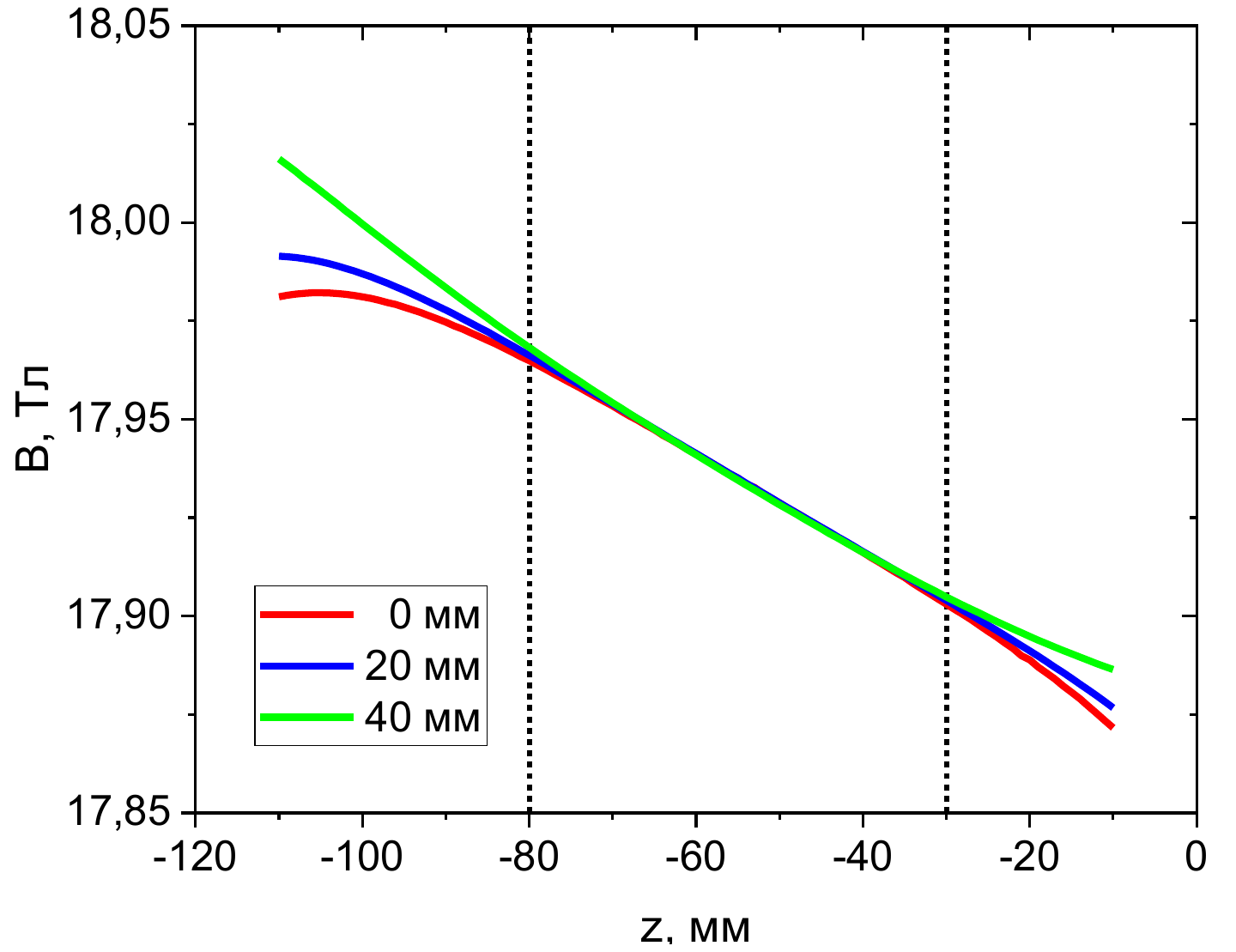}
    % \hfill
    \includegraphics[width=0.495\linewidth]{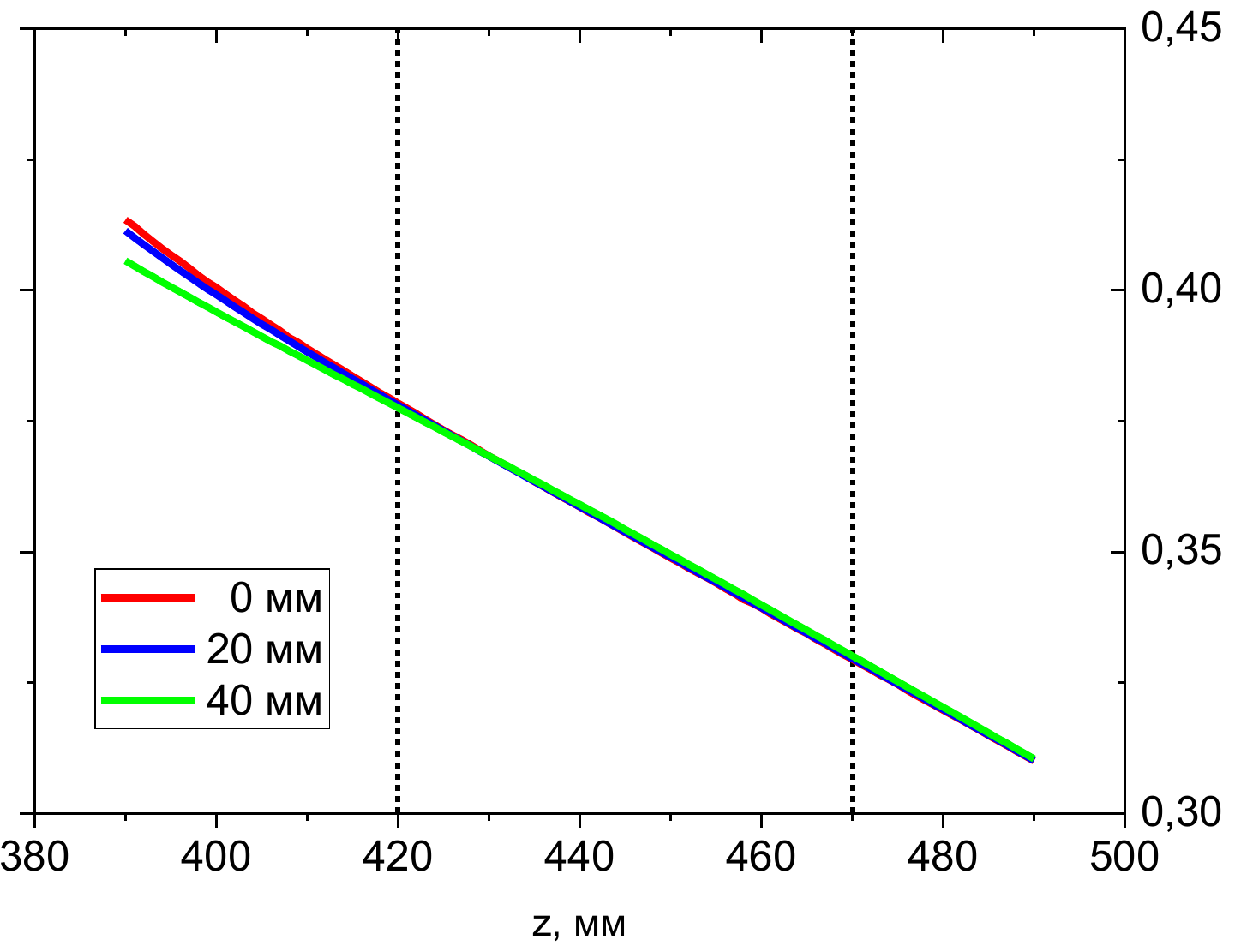}
    \hfill
    % Use \caption command for figure caption and label.
    \caption{Magnetic field profile as a function of distance from the center of the solenoid in two spin flip regions. The dotted lines indicate the boundaries of these regions.}\label{fig:fig5}
\end{figure}

\begin{figure}[!htb]%% placement specifier
    % Use \includegraphics command to insert graphic files. Place graphics files in working directory.
    \centering %% For centre alignment of image.
    \hfill
    \includegraphics[width=0.495\linewidth]{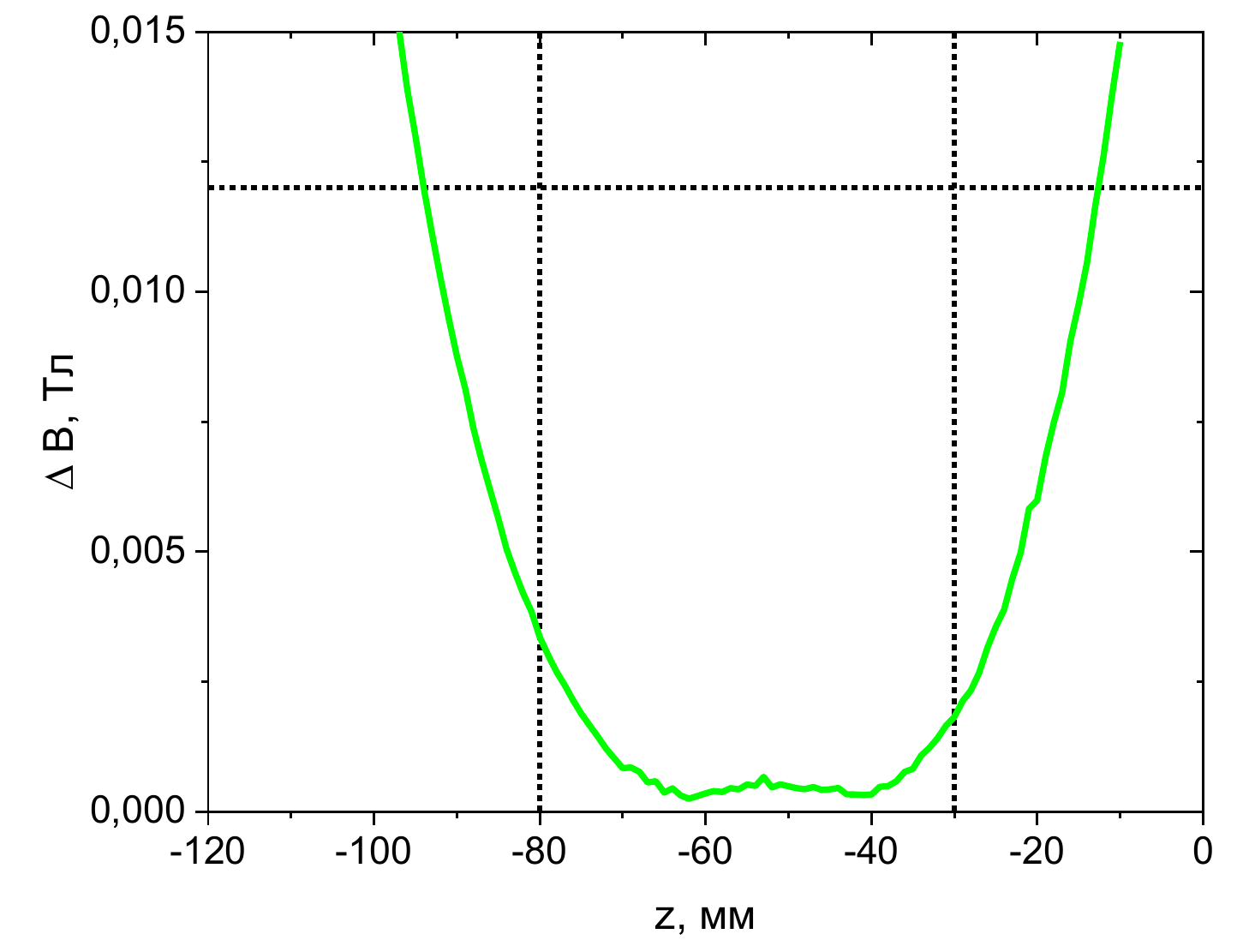}
    % \hfill
    \includegraphics[width=0.495\linewidth]{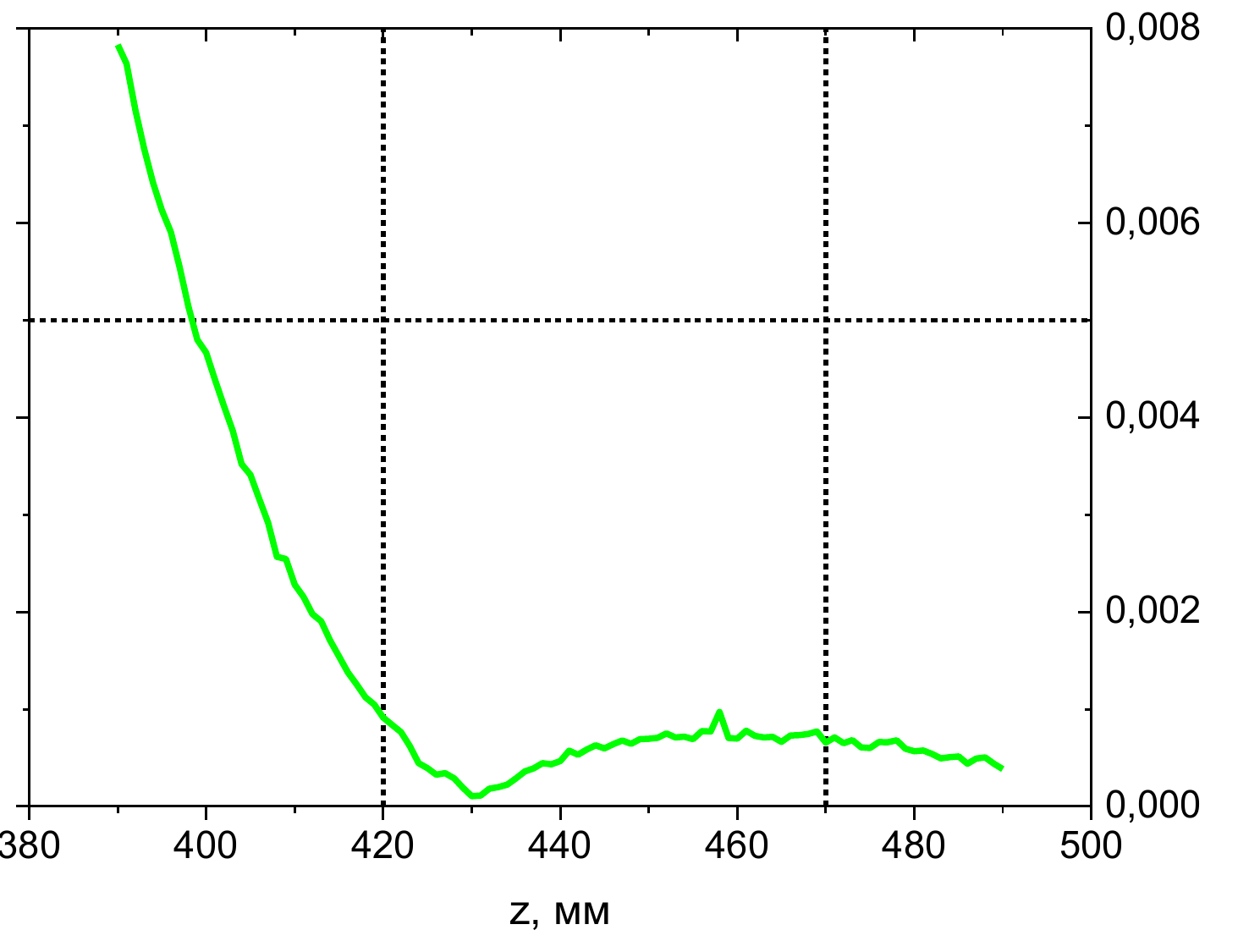}
    \hfill
    % Use \caption command for figure caption and label.
    \caption{The maximum deviation of the strength of the magnetic field from the field along the axis. Dotted lines indicate a rectangle that marks the maximum allowed deviation of the field from the field along the axis in the region of interest, which is the region where the spin flip occurs.}\label{fig:fig6}
\end{figure}

Figure~\ref{fig:fig5} shows the results of field strength calculations for two spin flip regions for several sections along the solenoid axis. In more detail, the difference in field strength at the maximum distance from the solenoid axis compared to the field strength on the axis is shown in Figure~\ref{fig:fig6}.

\begin{figure}[!htb]%% placement specifier
    % Use \includegraphics command to insert graphic files. Place graphics files in working directory.
    \centering %% For centre alignment of image.
    \hfill
    \includegraphics[width=0.495\linewidth]{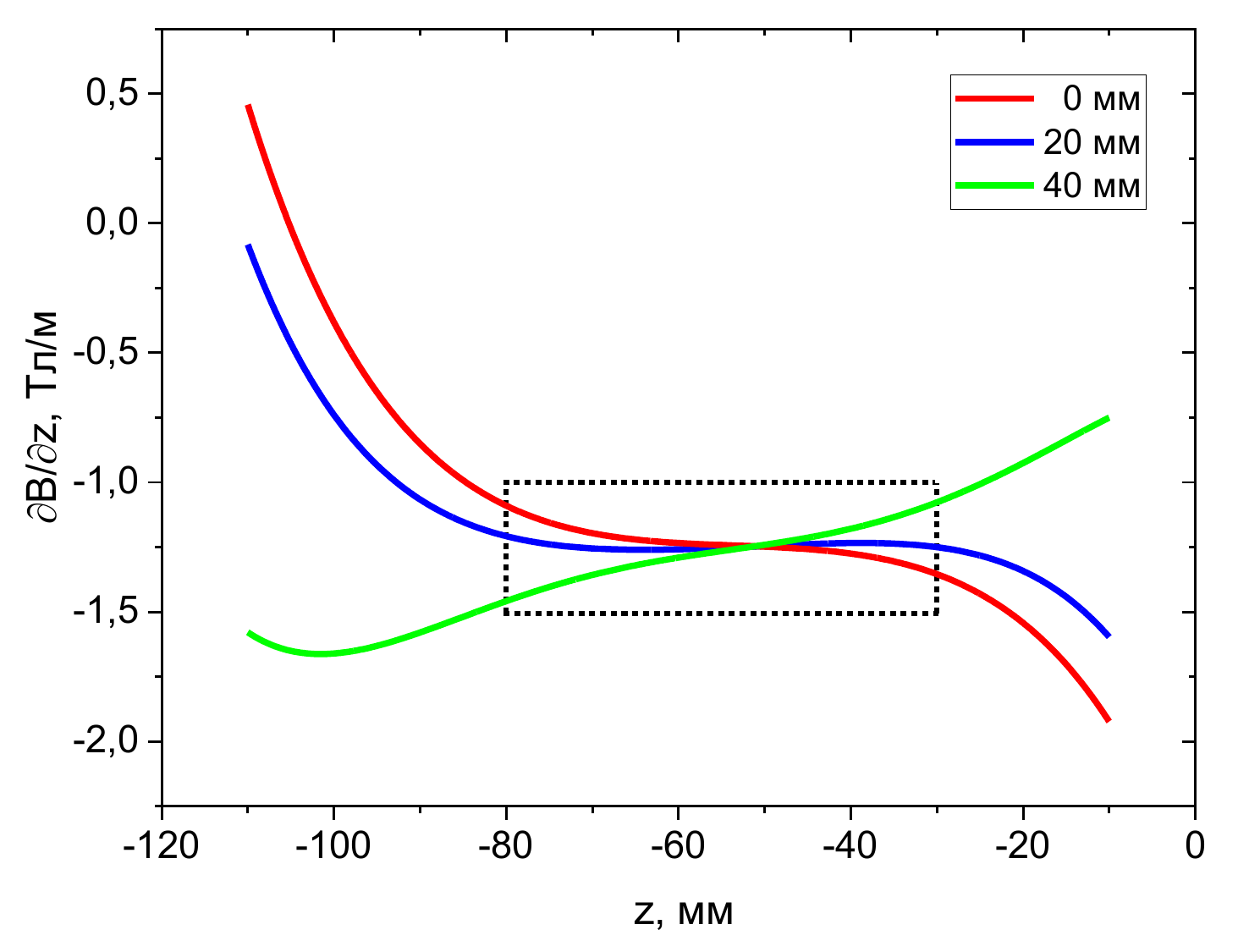}
    % \hfill
    \includegraphics[width=0.495\linewidth]{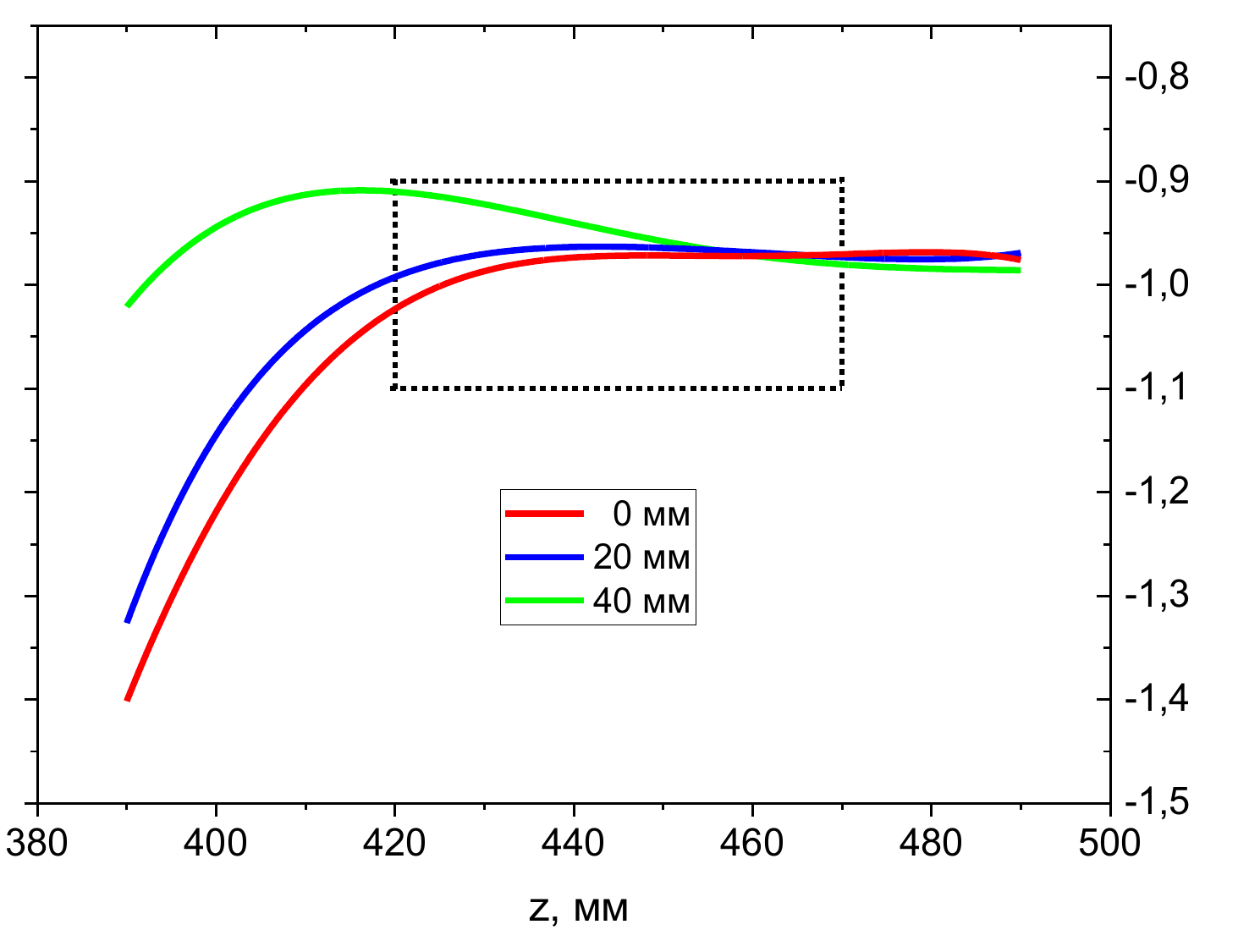}
    \hfill
    % Use \caption command for figure caption and label.
    \caption{The maximum deviation of the strength of the magnetic field from the field along the axis. Dotted lines indicate a rectangle that marks the maximum allowed deviation of the field from the field along the axis in the region of interest, which is the region where the spin flip occurs.}\label{fig:fig7}
\end{figure}

Figure~\ref{fig:fig7} shows the results of calculation of the magnetic field gradient in the spin flip regions. The square region bounded by the dotted lines defines the allowable range of the gradient and the region of spin flip.

As can be seen in the graphs in Figures~\ref{fig:fig5}-\ref{fig:fig7}, we were able to find a configuration for the magnetic system that creates a magnetic field that meets the requirements outlined in Section~\ref{sec:sec3} and ensures efficient neutron spin flip in both desired regions.

 \section{Estimates of the neutron deceleration time}
%% Labels are used to cross-reference an item using \ref command.
\label{sec:sec5}
The results of the calculations above give reason to expect high efficiency of the flipper itself. However, it should be noted that the idea of using a flipper with significant energy exchange with neutrons is based on the desire to produce a pulsed neutron flux at the flipper output. The main parameter that determines the gain in UCN density in the pulsed trap filling method is the ratio of the duration of a neutron pulse at the trap entrance to the period of repetition of these pulses. Therefore, determining the time required for neutron deceleration by the flipper and its dependence on neutron energy is crucial in designing a flipper-decelerator. Estimation of this deceleration time and its relationship with initial neutron velocity was the aim of the detailed calculations.

The Monte Carlo simulation of the motion of neutrons inside a guide, placed in the center of a solenoid, operated with a magnetic field map that was represented on a three-dimensional grid with a spacing of 1 mm. The magnetic field map was obtained through a separate calculation. For the sake of simplicity in calculation, the cross-section of the neutron guide was assumed to be rectangular, with dimensions of 80$\times$80 cm$^2$, and a value of 4 m/s was set as the boundary velocity for the neutron guide.

The transit time of the 1380 mm length magnetic field region has been calculated (see Figure~\ref{fig:fig4}), for neutrons whose velocity along the channel axis was in the range from 20.3 to 21.5 m/s. It has been assumed that the neutron spin change occurred in the center of the region, so that the neutron decelerates all the way in the magnetic field. The final velocities of the neutrons after passing through the flipper ranged from 3.5 to 5 m/s, to limit the range of deceleration times. The magnitude of the transverse velocity component remained practically unchanged and was limited by the boundary velocity of the neutron guide. Only neutrons with a full velocity of less than 6.9 m/s were considered in the calculation, as in a future UCN source, neutrons with higher velocities would not be able to be trapped.

\begin{figure}[!htb]%% placement specifier
%% Use \includegraphics command to insert graphic files. Place graphics files in working directory.
\centering%% For centre alignment of image.
\includegraphics[width=0.75\linewidth]{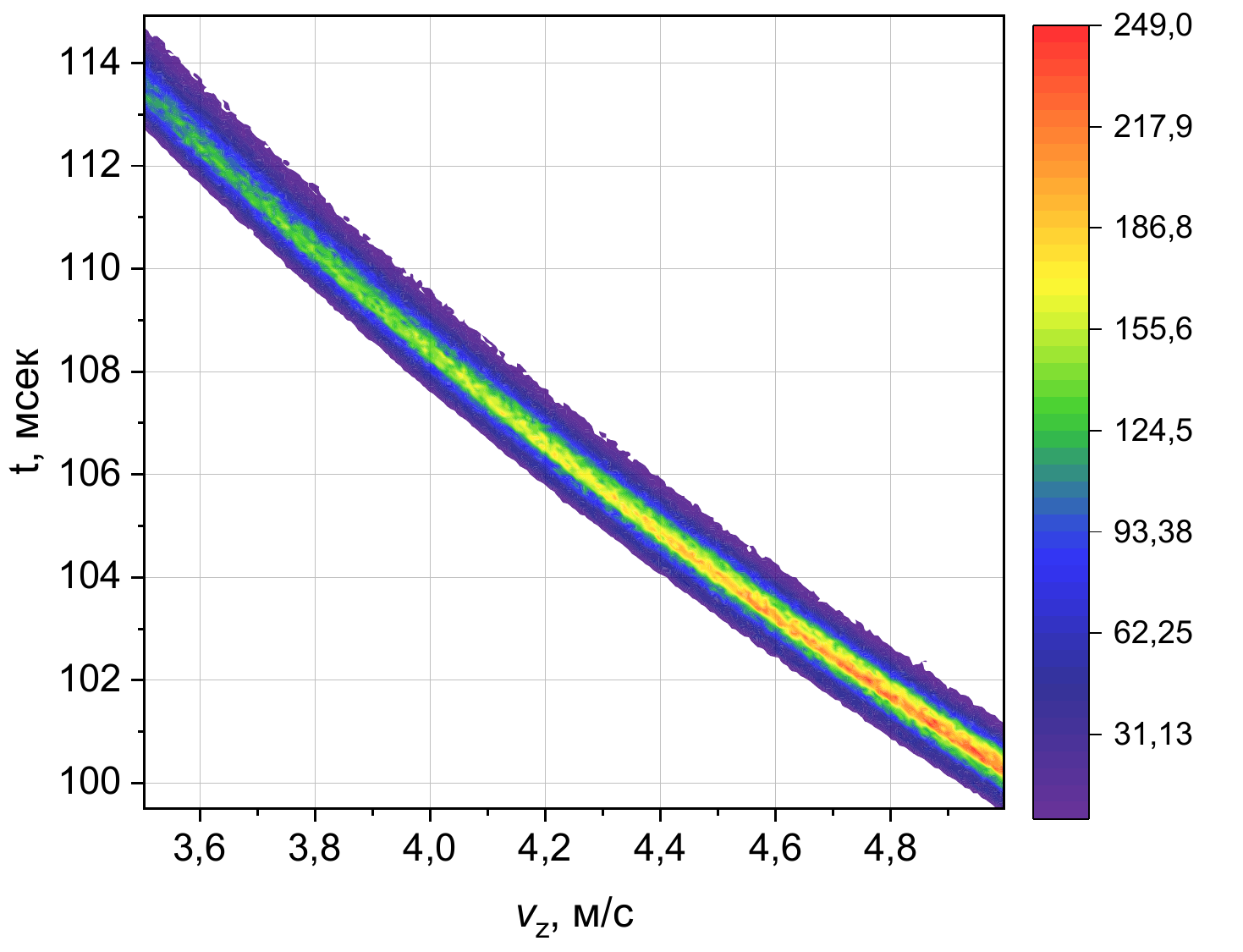}
%% Use \caption command for figure caption and label.
\caption{The flight time through a region with a magnetic field by neutrons, as a function of their final longitudinal velocity.}\label{fig:fig8}
\end{figure}

The calculation results are presented in Figure~\ref{fig:fig8}. The figure shows a map of the distribution of the number of neutrons by the flight time of a region with a magnetic field as a function of the final longitudinal velocity.  

\begin{figure}[!htb]%% placement specifier
%% Use \includegraphics command to insert graphic files. Place graphics files in working directory.
\centering%% For centre alignment of image.
\includegraphics[width=0.75\linewidth]{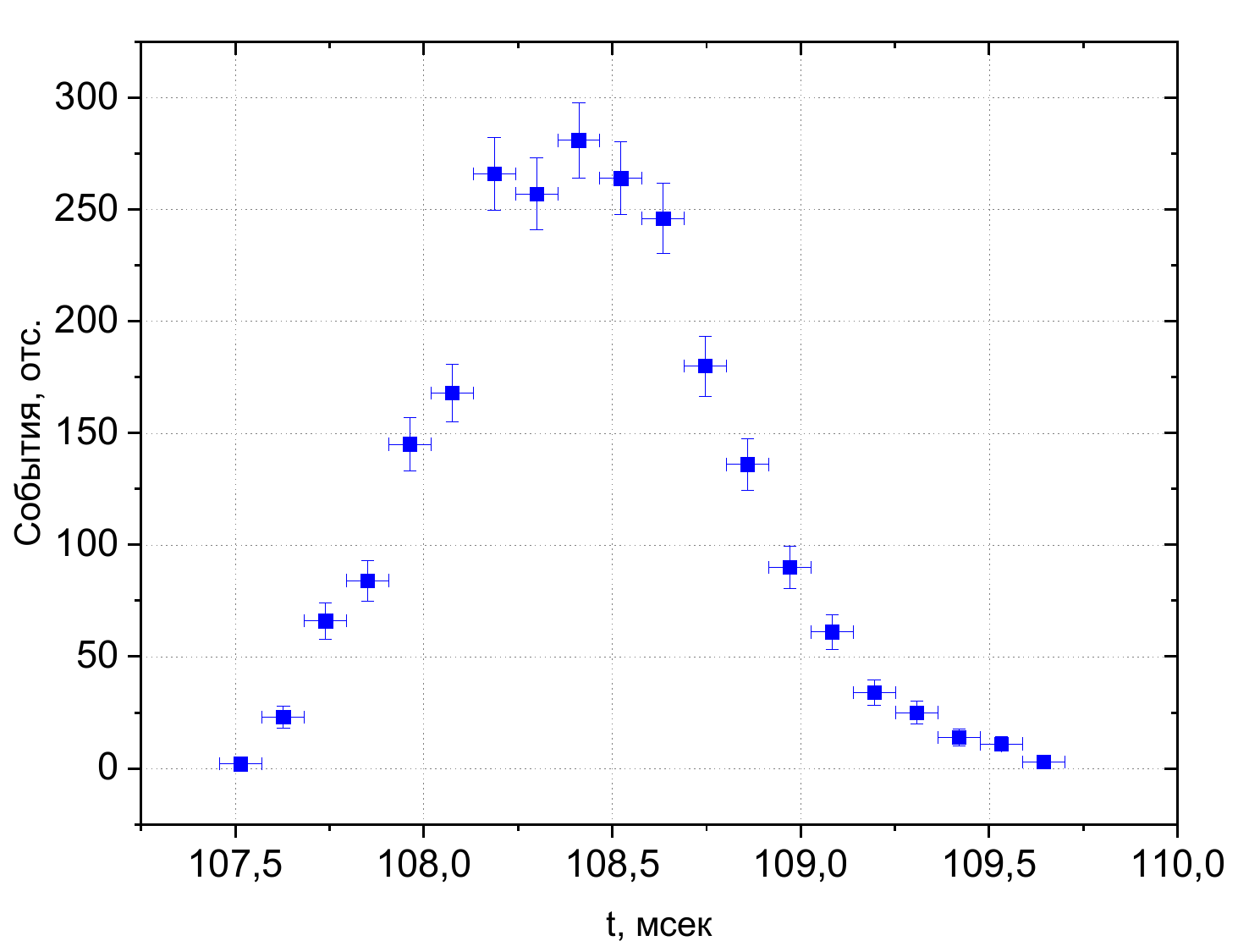}
%% Use \caption command for figure caption and label.
\caption{Deceleration time distribution for quasi-monochromatic neutrons with velocities of 4 m/s.}\label{fig:fig9}
\end{figure}

From this figure, it can be seen that the spread of deceleration times in the flipper is approximately 15 ms. Meanwhile, the deceleration time for monochromatic neutrons also has a certain distribution, which is probably due to the radial inhomogeneity of the magnetic field. Figure~\ref{fig:fig9} illustrates the distribution of flight times for neutrons with longitudinal velocities within a narrow range of 4 to 4.01 m/s. Note that the time dispersion of the order of one-and-a-half milliseconds is statistical in nature and, under these conditions, is presumably an unavoidable limit on the duration of the bunch of neutrons leaving the flipper. In contrast, a significantly greater variation in the duration of deceleration time as a function of neutron velocity is regular. Observe that neutrons with lowest velocities, which take the longest travel time from their birthplace to the flipper entrance, have the longest deceleration time. In \cite{FrankArXiv24}, it was found that there is a principal possibility to significantly compensate for an increase in the deceleration time, when the velocity decreases, by increasing the time for neutron transport from the source to the flipper \cite{FrankArXiv24}.

 \section{To the calculation of the RF resonator}
%% Labels are used to cross-reference an item using \ref command.
\label{sec:sec6}
Recall that for the gradient flipper to operate, it is necessary to have both a permanent magnetic field $B$ and an oscillating or rotating high-frequency magnetic field orthogonal to it, with a well-defined frequency $f=\gamma B/2\pi$ and amplitude, $B_1$, ensuring the fulfillment of the adiabatic condition \eqref{eq:eq2}. The selected value $B$ of the field and its gradient are consistent with the following parameters of high-frequency field: $f = 521.6$ MHz и $B_1 \geq 0.6$ mT. The choice of a solenoid as a potential source of a permanent magnetic field imposes certain constraints on the acceptable geometric parameters of the high-frequency resonator that generates the required RF field. The resonator should be positioned within the room temperature bore of the cryostat containing the solenoid and should cover the neutron guide through which the neutrons should propagate. A common type of resonator known as a "birdcage" may well satisfy these requirements. This type of resonator not only satisfies all geometric requirements, but also allows it to form a sufficiently uniform, rotating magnetic field with the required strength. A resonator of this type has already been used in an adiabatic flipper for UCN \cite{Holley12}, albeit with a substantially lower magnetic field strength of 1 T.

\begin{figure}[!htb]%% placement specifier
    % Use \includegraphics command to insert graphic files. Place graphics files in working directory.
    \centering %% For centre alignment of image.
    \hfill
    \includegraphics[ width=0.475\linewidth, height=.465\textwidth]{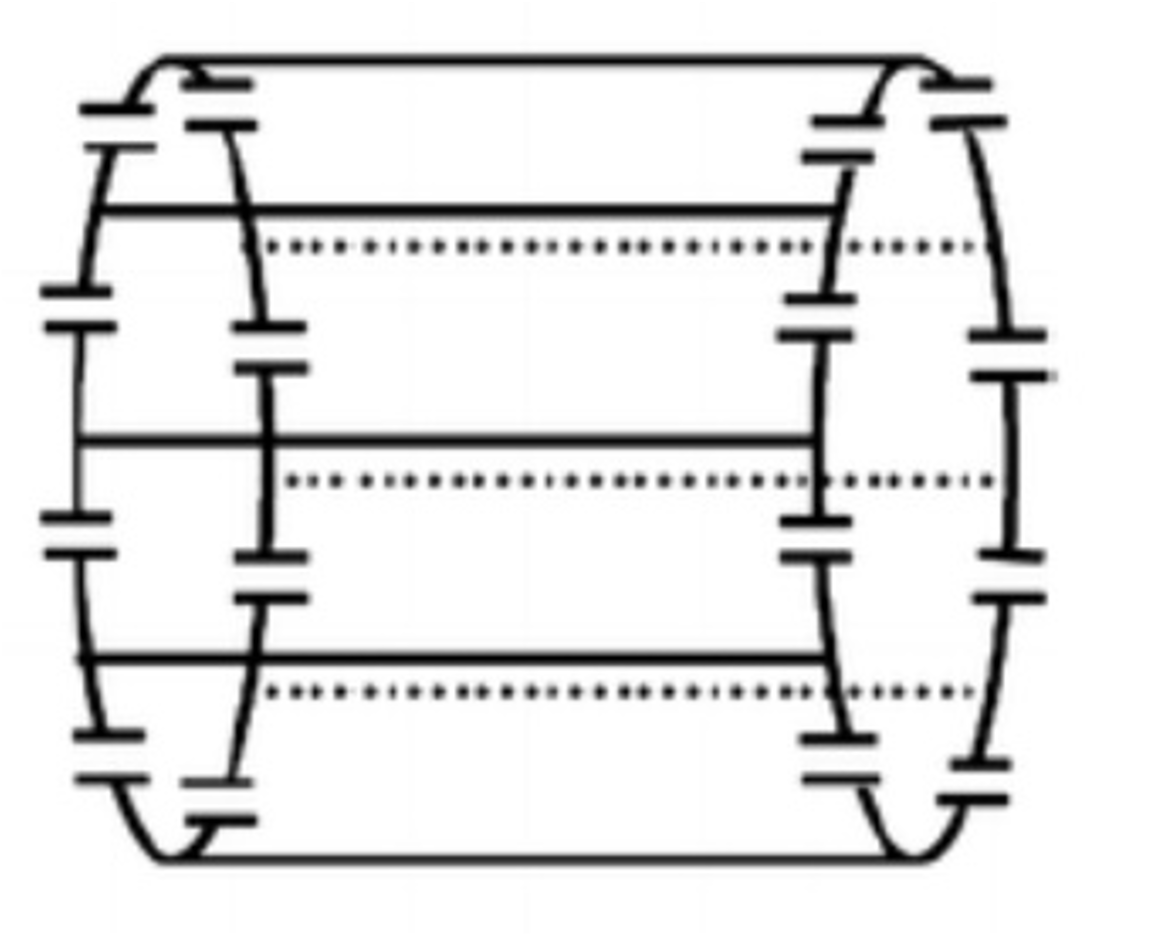}
    % \hfill
    \includegraphics[width=0.455\linewidth]{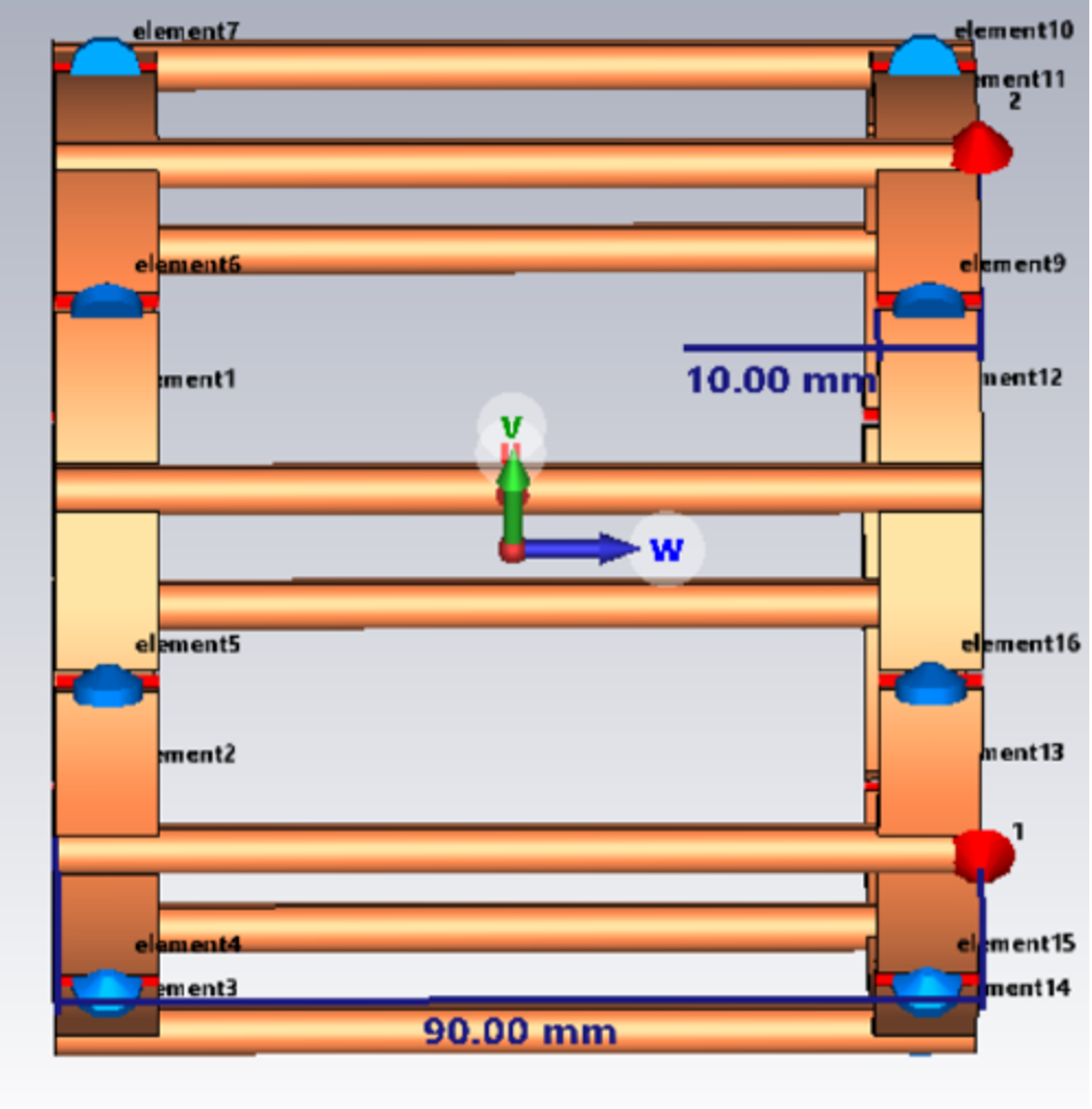}
    \hfill
    % Use \caption command for figure caption and label.
    \caption{Electrical circuit diagram for the resonator and its intended view.}\label{fig:fig10}
\end{figure}

Below, we will present some parameters of the proposed design for such a resonator. The chosen type of resonator structurally consists of two conducting rings connected by several metal legs. These rings have slots in which capacitors are inserted. The electrical circuit diagram for this system, as well as the intended view of the resonator, are shown in Figure~\ref{fig:fig10}.

\begin{figure}[!htb]%% placement specifier
   %% Use \includegraphics command to insert graphic files. Place graphics files in working directory.
\centering%% For centre alignment of image.
\includegraphics[width=0.95\linewidth]{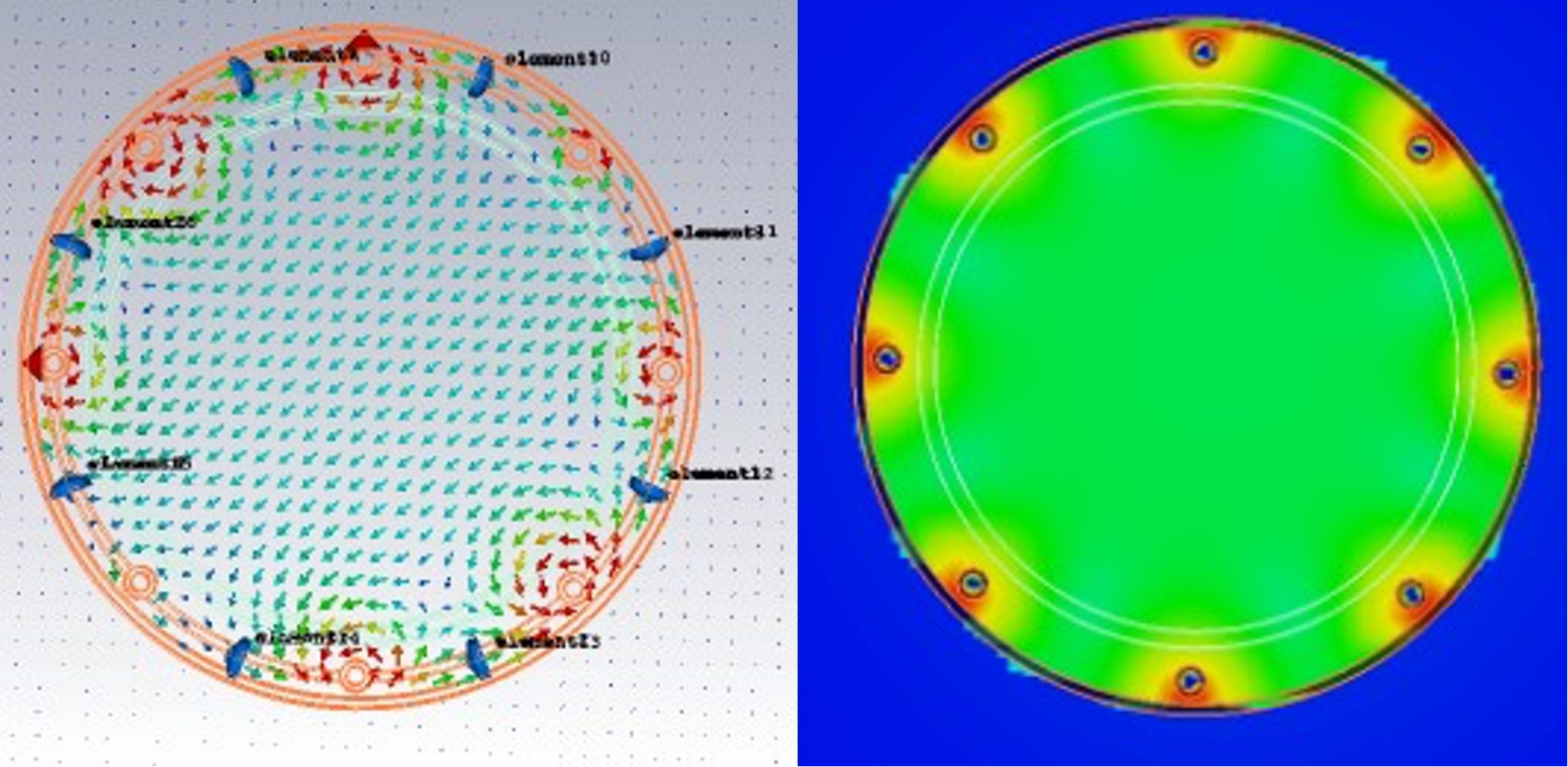}
%% Use \caption command for figure caption and label. 
\caption{The directions of the magnetic field vector in the cross-section of the resonator (left) and the pattern of magnetic energy distribution in it (right).}\label{fig:fig11}
\end{figure}

The electrical properties of the resonator have been calculated using the CST Studio Suite software package, following the procedure described in \cite{BernatPhD}. The outer diameter of the ring and the length of the proposed resonator are 10.2 mm and 9.0 cm, respectively. The number of legs is 8. With these chosen resonator parameters, it is possible to generate a rotating magnetic field of sufficiently high uniformity within the resonator (see Figure~\ref{fig:fig11}).

The resonator of the main flipper must generate an almost uniform rotating magnetic field with the required frequency and amplitude, at a power input of approximately 5 kW. The device is assumed to operate in a pulse mode, turning on only for the duration of a neutron bunch passage, which is on the order of 15–20 ms, and with a repetition frequency of 5 Hz. This reduces the average power input by approximately one order of magnitude.

Regarding the resonator for the flipper-valve, its design is similar to that described above, with the exception that the frequency of the RF field should be 11.5 MHz. Since the power consumed by the resonator is proportional to the fourth power of frequency, this resonator has a relatively low power requirement. It should be noted that, in accordance with its intended function, the flip-valve must also operate in pulse mode.

 \section{Conclusion}
%% Labels are used to cross-reference an item using \ref command.
\label{sec:conc}
The work is devoted to the development of a conceptual design for a gradient spin flipper — neutron decelerator, which is the main component of a designed UCN source for a pulsed reactor. In close cooperation between the JINR group and SuperOx, a preliminary design of a stationary gradient magnet for the adiabatic spin flipper has been developed. A thorough calculation of the magnetic field configuration has been performed. The movement of neutrons in the magnetic field generated by the designed magnetic system has been simulated, and the deceleration time of neutrons in the spin flipper has been analyzed. For the selected range of UCN velocities generated by the device, the main contribution to the spread in deceleration times is the dependence of the latter on the initial neutron velocity. Since the dependence of deceleration time on initial neutron velocity is rather regular, there is hope for the possibility to significantly compensate for an increase in the deceleration time, when the velocity decreases, by increasing the time for neutron transport from the source to the flipper \cite{FrankArXiv24}. While it is evidently unavoidable to have some residual dispersion in the deceleration times caused by the radial inhomogeneity of the magnetic field in this design, this does lead to a slight extension of the duration of the neutron bunches, compared to the period of their repetition. Although the residual dispersion of the deceleration time caused by the radial inhomogeneity of the magnetic field in this design is apparently unavoidable, it leads to a slight increase in the duration of neutron bunches compared to the period of their repetition.

The findings of this work have significant practical implication for the development of a new UCN source. The results obtained give grounds for hope that the idea of creating a UCN source based on pulsed accumulation in a trap using non-stationary neutron deceleration is feasible.

\section*{Acknowledgements}
The authors would like to express their gratitude   to V.N. Shvetsov and A.N. Chernikov for support and useful discussions. 

\section*{Funding}

The work was carried out in accordance with the topic 01-3-1146-3-2024/2028 of the JINR topical plan.

\section*{Conflicts of Interest}

The authors declare no conflicts of interest.


%% References
%%   \cite{label}         ==>>  [#]
%%   \cite[chap. 2]{label} ==>> [#, chap. 2]

%% References with BibTeX:
%\bibliographystyle{nsr}
%\bibliography{sample}

\begin{thebibliography}{20}
\expandafter\ifx\csname url\endcsname\relax
  \def\url#1{\texttt{#1}}\fi
\expandafter\ifx\csname urlprefix\endcsname\relax\def\urlprefix{URL }\fi
\expandafter\ifx\csname href\endcsname\relax
  \def\href#1#2{#2} \def\path#1{#1}\fi

\bibitem{Ananiev77}
V.~D. Ananiev, D.~I. Blokhintsev, Y.~M. Bulkin, et~al., {IBR}-2 - pulsed
  reactor for neutron investigations, Pribory i Tekhnika Ehksperimenta 5 (1977)
  17--35, in {R}ussian.

\bibitem{Antonov69}
A.~V. Antonov, A.~I. Isakov, M.~V. Kazarnovskii, V.~E. Solodilov, Concerning a
  gas of ultracold neutrons in a trap, Journal of Experimental and Theoretical
  Physics Letters 10 (1969) 241--244.
  
  \bibitem{Shapiro71}
F.~L. Shapiro, Remarks on the measurement of phases of structural amplitudes in
  neutron diffraction and on the accumulation of neutrons, Physics of
  Elementary Particles and Atomic Nuclei 2~(4) (1971) 975--979, in {R}ussian.

\bibitem{Shapiro74}
F.~L. Shapiro, {Ultracold
  Neutrons}, Springer US, Boston, MA, 1974, pp. 259--284.
\newblock \href {https://doi.org/10.1007/978-1-4613-4499-5_10}
  {\path{doi:10.1007/978-1-4613-4499-5_10}}.

\bibitem{Frank00}
A.~I. Frank, R.~Gahler, Time focusing of neutrons, Physics of Atomic Nuclei 63
  (2000) 545--547.

\bibitem{Arimoto12}
Y.~Arimoto, P.~Gertenbort, S.~Imajo, et~al., Demonstration of focusing by a
  neutron accelerator, Physical Review A 86 (2012) 023843.
  
 \bibitem{Frank22}
A.~I. Frank, G.~V. Kulin, N.~V. Rebrova, M.~A. Zakharov, On the {P}ossibility
  of {C}reating a {UCN} {S}ource at a {P}eriodic {P}ulsed {R}eactor, Physics of
  Particles and Nuclei 53~(1) (2022) 33--44.

\bibitem{Nesvizhevsky22}
N.~N. Nesvizhevsky, A.~O. Sidorin, Production of {U}ltracold {N}eutrons in an
  {E}scaping {D}ecelerating {T}rap, Physics of Particles and Nuclei Letters 19
  (2022) 162--175.
  
 \bibitem{FrankPEPANL23}
A.~I. Frank, G.~V. Kulin, M.~A. Zakharov,
  {On a {N}ew {P}ossibility of
  {P}ulsed {A}ccumulation of {U}ltra {C}old {N}eutrons in a {T}rap}, Physics of
  Particles and Nuclei Letters 20~(4) (2023) 664--667.
\newblock \href {https://doi.org/10.1134/S1547477123040295}
  {\path{doi:10.1134/S1547477123040295}}.


\bibitem{FrankArXiv24}
A.~I. Frank, G.~V. Kulin, M.~A. Zakharov, et~al.,
 {To the {UCN} source with pulsed filling of a trap}, arXiv:2412.06460 [physics.ins-det] (2024).
\newblock \href {https://doi.org/10.48550/arXiv.2412.06460}
  {\path{doi:10.48550/arXiv.2412.06460}}.

\bibitem{Drabkin60}
G.~M. Drabkin, R.~A. Zhitnikov, Production of \enquote{supercold} polarized
  neutrons, Soviet Journal of Experimental and Theoretical Physics 11~(3)
  (1960) 729--730.
  
\bibitem{Kruger80}
E.~Kruger, Acceleration of polarized neutrons by rotating magnetic field, Nucleonika 25
  (1980) 889--893.
  
\bibitem{Alefeld81}
B.~Alefeld, G.~Badurek, H.~Rauch,
  {Observation of the neutron magnetic
  resonance energy shift}, Zeitschrift f\"ur Physik B Condensed Matter 41~(3)
  (1981) 231--235.
\newblock \href {https://doi.org/10.1007/BF01294428}
  {\path{doi:10.1007/BF01294428}}.

\bibitem{Egorov74}
A.~I. Egorov, V.~M. Lobashov, V.~A. Nazarenko, \text{et al}, Production,
  storage, and polarization of ultracold neutrons, Soviet Journal of Nuclear
  Physics 19~(2) (1974) 147--152.
  
\bibitem{Luschikov84}
V.~I. Luschikov, Y.~V. Taran, On the calculation of the neutron adiabatic
  spin-flipper, Nuclear Instruments and Methods in Physics Research 228 (1984)
  159--160.
  
\bibitem{Grigoriev97}
S.~V. Grigoriev, A.~Okorokov, V.~Runov, Peculiarities of the construction and
  application of a broadband adiabatic flipper of cold neutrons, Nuclear
  Instruments and Methods in Physics Research Section A: Accelerators,
  Spectrometers, Detectors and Associated Equipment 384 (1997) 451--456.
  
\bibitem{Holley12}
A.~T. Holley, L.~J. Broussard, J.~L. Davis, et~al., {A high-field adiabatic
  fast passage ultracold neutron spin flipper for the {UCNA} experiment},
  Review of Scientific Instruments 83~(7) (2012) 073505.
\newblock \href {https://doi.org/10.1063/1.4732822}
  {\path{doi:10.1063/1.4732822}}.
  
\bibitem{Weinfurter88}
H.~Weinfurter, G.~Badurek, H.~Rauch, D.~Schwahn,
 {Inelastic action of a gradient
  radio-frequency neutron spin flipper}, Zeitschrift f\"ur Physik B Condensed
  Matter 72~(2) (1988) 195--201.
\newblock \href {https://doi.org/10.1007/BF01312135}
  {\path{doi:10.1007/BF01312135}}.

\bibitem{Molodyk21}
A.~Molodyk, S.~Samoilenkov, A.~Markelov, et~al.,
 {Development and large volume production of extremely high current density $\rm YBa_2Cu_3O_7$ superconducting wires for fusion}, Scientific Reports 11 (2021) 2084.
\newblock \href {https://doi.org/10.1038/s41598-021-81559-z}
  {\path{doi:10.1038/s41598-021-81559-z}}.

\bibitem{Baburin24}
K.~Baburin, E.~Zapretilina, N.~Hitrov, et~al.,
{Design, Manufacturing and Tests of an All-HTS 20 T Magnet}, in: IEEE Transactions on Applied Superconductivity vol. 34, no. 5, Art no. 4605604, 2024, pp. 1--4.
\newblock \href {https://doi.org/10.1109/TASC.2024.3391755}
  {\path{doi:10.1038/s41598-021-81559-z}}.

\bibitem{BernatPhD}
P.~T. Bernat, {Design and {S}imulation of a {B}irdcage {C}oil using {CST} {S}tudio {S}uite for {A}pplication at 7{T}}, PhD thesis, UPC, Escola T‘ecnica Superior d’Enginyeria de Telecomunicaci o de Barcelona, Departament de Teoria del Senyal i Comunicacions, Feb 2013, \url{http://hdl.handle.net/2099.1/18371}
 

\end{thebibliography}
%% Add your reference list in the sample.bib file.
%% For collaboration in authors put the collaboration in {}, e.g. {SPD Collaboration}

%\begin{thebibliography}{00}
%%
%% For numbered reference style
%% Text of bibliographic item example
%\bibitem{einstein}
% A. Einstein,
% Zur Elektrodynamik bewegter K{\"o}rper 
% (On the electrodynamics of moving bodies),
% Annalen der Physik
% 322 (10) (1905)
% 891–921, (in German)
% \href{http://dx.doi.org/10.1002/andp.19053221004}{doi:10.1002/andp.19053221004}.

%\end{thebibliography}

\end{document}